 \documentclass[twocolumn]{aastex61}

\shorttitle{Light Curve Analysis of V959~Mon}
\shortauthors{Hachisu \& Kato}

\begin{document}

\title{A Light Curve Analysis of Gamma-ray Nova V959~Mon 
-- Distance and White Dwarf Mass}


\author{Izumi Hachisu}
\affil{Department of Earth Science and Astronomy, 
College of Arts and Sciences, The University of Tokyo,
3-8-1 Komaba, Meguro-ku, Tokyo 153-8902, Japan} 
\email{hachisu@ea.c.u-tokyo.ac.jp}


\author{Mariko Kato}
\affil{Department of Astronomy, Keio University, 
Hiyoshi, Kouhoku-ku, Yokohama 223-8521, Japan} 

\begin{abstract}
V959~Mon is one of the gamma-ray detected novae. 
It was optically discovered about 50 days after the gamma-ray detection
due to proximity to the Sun.  The nova speed class is unknown because of 
lack of the earliest half of optical light curve and short supersoft 
X-ray phase due to eclipse by the disk rim.  Using the universal 
decline law and time-stretching method, we analyzed the data of V959~Mon 
and obtained nova parameters.  We estimated the distance modulus
in the $V$ band to be $(m-M)_V=13.15\pm0.3$ for the reddening of
$E(B-V)=0.38\pm0.01$ by directly comparing with the similar type of novae, 
LV~Vul, V1668~Cyg, IV~Cep, and V1065~Cen.  The distance to V959~Mon
is $2.5\pm0.5$~kpc.  If we assume that the early phase light curve of
V959~Mon is the same as that of time-stretched light curves of
LV~Vul, our model light curve fitting suggests 
that the white dwarf (WD) mass is $0.9-1.15~M_\sun$, being consistent
with a neon nova identification. 
At the time of gamma-ray detection the photosphere of nova envelope
extends to $5-8~R_\sun$ (about two or three times the binary separation)
and the wind mass-loss rate is $(3-4)\times 10^{-5} M_\sun$~yr$^{-1}$.  
The period of hard X-ray emission is consistent with the time of appearance 
of the companion star from the nova envelope.  The short supersoft X-ray
turnoff time is consistent with the epoch when the WD photosphere shrank
to behind the elevating disk rim, that occurs 500 days before nuclear
burning turned off.
\end{abstract}





\section{Introduction}
\label{introduction}
Recently, GeV gamma-rays have been detected in 
several classical and symbiotic novae with 
the {\it Fermi}/Large Area Telescope (LAT) \citep[e.g.,][]{ack14}. 
These gamma-ray novae show wide varieties in the speed class (fast and
slow), nova type (CO and ONe), and companion type (a close binary with
a red dwarf companion and a wide binary with a red giant companion). 
In some gamma-ray novae hard X-ray emission was also detected. 
Gamma-ray could be produced in shock interaction between ejecta
and companion or circumbinary matter, or collision between ejecta shells
(internal shock).  The exact origin of gamma-ray and relation with 
hard X-ray emission in the individual objects are not known. 

Although gamma-rays were detected in various types of novae, 
the gamma-ray properties appear similar to one another. Therefore, 
it has been argued that there is a common mechanism of gamma-ray emission 
independent of nova type, thus, all novae are potential gamma-ray emitters 
\citep[e.g.,][]{ack14}.  If this is the case, the gamma-ray detection is 
due to close proximity and non-detected novae should be more distant.  
\citet{ack14} concluded that all the {\it Fermi}/LAT-detected novae 
have estimated distances of $\lesssim 4$ to 5 kpc,  
i.e., V407~Cyg (2.7~kpc), V1324~Sco (4.5~kpc), V959~Mon (3.6~kpc),
and V339~Del (4.2~kpc).  

The distances of novae, however, are always debated. 
For example, \citet{fin15} derived the distance of V1324~Sco to be 
$d > 6.5$~kpc based on their estimated reddening of $E(B-V)=1.16\pm0.12$.
This is much larger than the above distance, $d=4.5$~kpc, 
that Ackerman et al. derived from various Maximum Magnitude vs.
Rate of Decline (MMRD) relations. 
As well known, the MMRD relations are statistical relations and not a good
indicator for individual novae \citep[see, e.g.,][]{dow00, kas11, shara17}.
For V407~Cyg, \citet{hac18k} recently redetermined the distance to be
$d=3.9$~kpc from the revised period-luminosity
relation of Miras \citep{ita11}. This value is larger than $d=2.7$~kpc  
that Ackerman et al. adopted.  To elucidate the nature of gamma-ray
detected novae, we need an accurate distance of each nova. 
One of the aims of this work is to obtain the distance of V959~Mon. 

The classical nova V959~Mon was optically discovered by S. Fujikawa
on UT 2012 August 9.8 (JD~2456149.3) at mag 9.4 \citep{fuj12}. 
Due to solar conjunction, the nova already entered 
the nebular decline phase when it was discovered. 
The optical peak was possibly substantially (more than 50 days) 
before the discovery \citep[e.g.,][]{mun13b}.  \citet{grei12} identified
the progenitor of V959~Mon using images from the INT Photometric 
H-Alpha Survey (IPHAS) at $r\sim17.9$ and $i\sim17.2$ mag.

Fermi J0639+0548 is a gamma-ray source detected on UT 2012 June 22
(JD~2456100.5) \citep{ack14}, 48 days before the optical discovery
of V959~Mon.  \citet{che12} reported that the position of V959~Mon
is consistent with that of the gamma-ray object Fermi J0639+0548
and the spectra of V959~Mon show striking resemblance
to that of the fast ONe nova V382 Vel 1999 well after optical peak.
\citet{mun13a} suggested that V959~Mon is a neon nova
because of very strong lines of [\ion{Ne}{3}], [\ion{Ne}{4}],
and [\ion{Ne}{5}] in the spectrum obtained on UT 2013 January 5.938
(JD~2456298.438).
Thus, V959~Mon was identified as a gamma-ray nova and also as a neon nova.
%

V959~Mon entered the supersoft X-ray source (SSS) phase 
on UT 2012 November 18 (JD~2456249.5) \citep{nel12b}.
The UV and X-ray emission observed with {\it Swift} showed 
modulation that indicates 7.1~hr orbital period \citep{osb13}.
\citet{osb13} argued that the orbital variation
of the X-ray light curve is caused by a partial eclipse of extended emission
by an accretion disk rim which is raised at the point of impact
of the stream from the secondary star \citep[see also][]{pag13}.

\citet{mun13b} presented the detailed photometric and spectroscopic data
of V959~Mon and showed the orbital light curves of the 7.1~hr period.
\citet{rib13} presented a model of the morphology of the ejected shell
to reproduce line profile spectra of the [\ion{O}{3}] 4959, 5007 and
derived a probable orbital inclination angle of $i=82\arcdeg\pm6\arcdeg$.

\begin{figure}
\epsscale{1.2}
\plotone{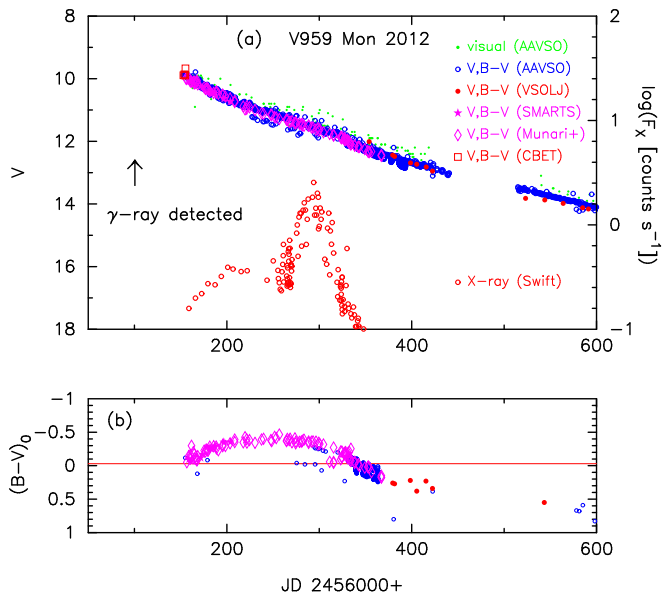}
\caption{
Optical light curve, X-ray, and color evolutions
of V959~Mon on a linear timescale.
(a) The visual data are taken from AAVSO (green dots).
The $V$ data are taken from AAVSO (open blue circles),
VSOLJ (filled red circles), SMARTS (filled magenta stars) \citep{wal12}, 
\citet{mun13b} (open magenta diamonds), and CBET No.3202 (open red squares).
We add {\it Swift}/XRT data (0.3-10 keV, open red circles) taken
from the {\it Swift} web site \citep{eva09}.
Gamma-rays were first detected on JD~2456100.5 at the epoch denoted
by the black arrow.
(b) The $(B-V)_0$ are dereddened with $E(B-V)=0.38$.
The horizontal solid red line, $(B-V)_0=-0.03$, indicates
the color of optically thick free-free emission.
\label{v959_mon_v_bv_ub_color_curve}}
\end{figure}

Figure \ref{v959_mon_v_bv_ub_color_curve}(a) summarizes the visual, $V$, and 
X-ray light curves of V959~Mon, and 
Figure \ref{v959_mon_v_bv_ub_color_curve}(b) shows $(B-V)_0$, which 
are dereddened with $E(B-V)=0.38$ after \citet{mun13b}.  
The lack of optical peak and following light curve data makes
this nova difficult to be categorized into a particular speed class. 
Moreover, the supersoft X-ray phase is interrupted in the mid way,
possibly due to occultation by the disk rim, which prevents the WD mass
from being estimated by the SSS duration \citep[e.g.,][]{hac10k}.  

In the present work, we analyze the data in Figure 
\ref{v959_mon_v_bv_ub_color_curve}, in the context of our nova light curve
analysis for more than 60 objects \citep[e.g.,][]{hac06kb, hac10k, hac14k,
hac15k, hac16k, hac16kb, hac18k}, and derive outburst parameters of 
V959~Mon as much as possible.  Our paper is organized as follows.
First we describe the time-stretching method of nova light curves,
and derive the distance modulus in the $V$ band,
$(m-M)_V$, toward V959~Mon in Section \ref{v959_mon}.
In Section \ref{color-magnitude_diagram}, we show similarity in 
the color-magnitude diagram between V959~Mon and LV~Vul (as well as
IV~Cep and V1065~Cen) and confirm that our obtained values of $(m-M)_V$
and $E(B-V)$ are reasonable.  
Section \ref{model_light_curve} describes our model
light curves that fit with the $V$ and X-ray data of V959~Mon.
We derive the WD mass, $M_{\rm WD}$, and present models for hard X-ray
and short duration of the SSS phase.
We discuss the distance modulus $(m-M)_V$, extinction 
$E(B-V)$, and distance $d$ in Section \ref{discussion}.  
Conclusions follow in Section \ref{conclusions}.


\begin{figure}
\epsscale{1.25}
\plotone{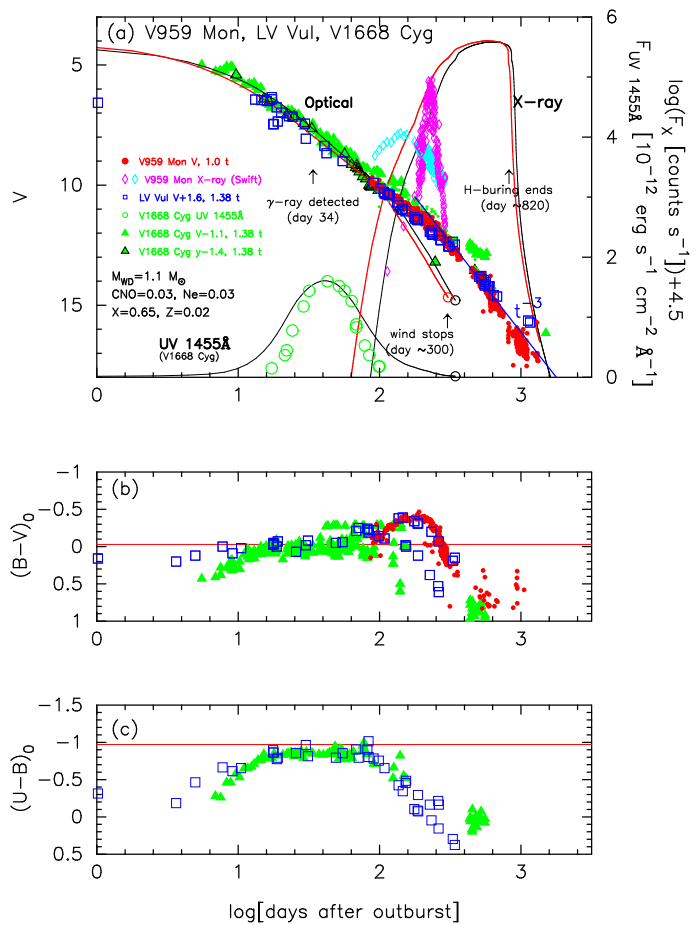}
\caption{
The optical light curves and color evolutions of V959~Mon (filled red
circles) on a logarithmic timescale as well as LV~Vul and V1668~Cyg.
The timescales of LV~Vul and V1668~Cyg are stretched by a factor of 1.38.
The data of V959~Mon are the same as those in Figure 
\ref{v959_mon_v_bv_ub_color_curve}.
The LV~Vul and V1668~Cyg data are taken from Figures 4 and 1
of \citet{hac16kb}, respectively.
In panel (a), we plot the model $V$ (blackbody plus free-free)
and X-ray (blackbody) light curves of a $1.1~M_\sun$ WD
(red lines) with the envelope chemical composition
of Ne nova 3 \citep{hac16k}, taking $(m-M)_V=13.15$ for V959~Mon.
Another set of model light curves of $V$, UV~1455\AA,
and X-ray (black lines) are those of a $0.98~M_\sun$ WD 
with the chemical composition of CO nova 3 for V1668~Cyg \citep{hac16k}.
The detection of gamma-ray on UT 2012 June 22 was indicated
by the black arrow, which corresponds to day 34 in this figure.
We depict the hard (0.8-10~keV: open cyan diamonds) 
and soft (0.3-0.8~keV: open magenta diamonds) X-ray fluxes separately,
both of which are taken from \citet{pag13}.
The open circles at the right edge of each $V$ model light curve correspond
to the epoch when optically thick winds stop.
We also add the free-free emission
decay trend along the $F_\nu \propto t^{-3}$ line 
after optically thick winds stop.  
In panel (b), we dereddened the colors with $E(B-V)=0.38$.
The horizontal solid red lines,  (b) $(B-V)_0=-0.03$
and (c) $(U-B)_0=-0.97$, indicate the colors of optically 
thick free-free emission, respectively.
See the text for detail.
\label{v959_mon_v1668_cyg_lv_vul_v_bv_ub_logscale_no3}}
\end{figure}


\begin{figure}
\epsscale{1.2}
\plotone{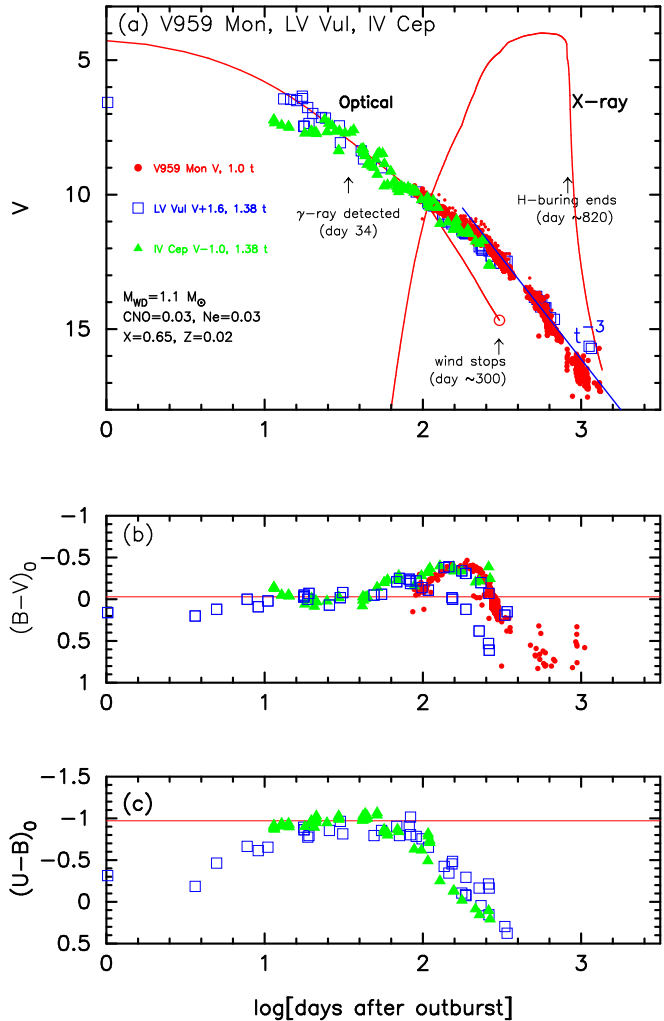}
\caption{
Same as Figure
\ref{v959_mon_v1668_cyg_lv_vul_v_bv_ub_logscale_no3}, but we compare
with IV~Cep 1971 instead of V1668~Cyg.
The timescale of IV~Cep is the same as that of LV~Vul, 
$f_{\rm s}=1.0$ against LV~Vul.
The data of IV~Cep are the same as those in Figure 21 of \citet{hac16kb};
the original data are taken from \citet{mac72} and \citet{koh73}.
In panels (b) and (c), we dereddened the colors of IV~Cep with $E(B-V)=0.65$.
See the text for detail.
\label{v959_mon_lv_vul_iv_cep_v_bv_ub_logscale_no3}}
\end{figure}


\begin{figure}
\epsscale{1.2}
\plotone{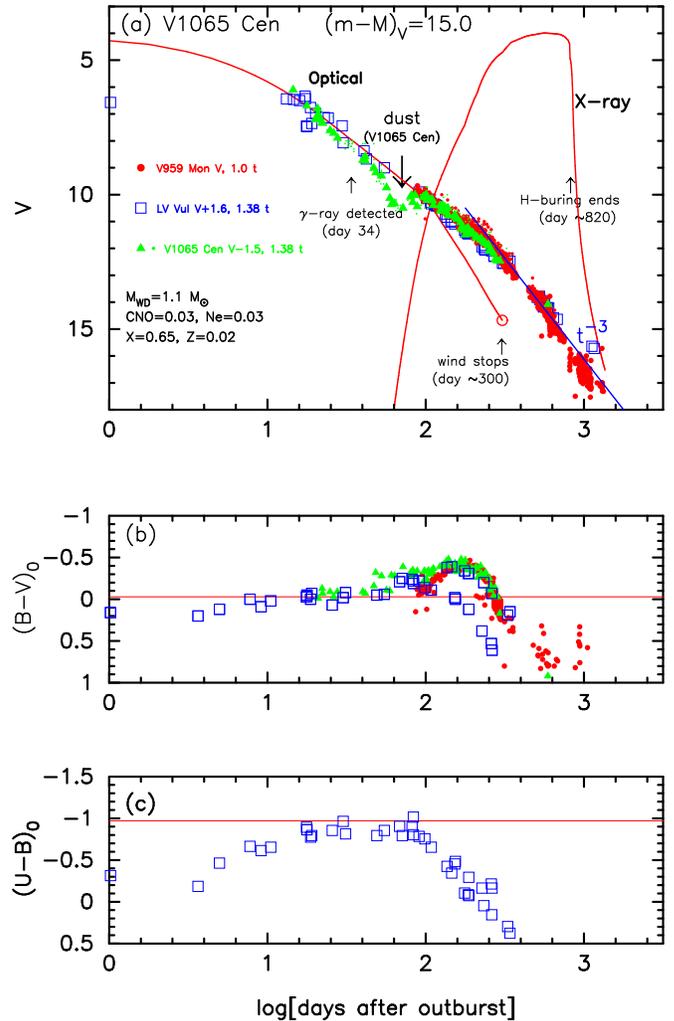}
\caption{
Same as Figure
\ref{v959_mon_v1668_cyg_lv_vul_v_bv_ub_logscale_no3}, but we compare
with V1065~Cen 2007 instead of V1668~Cyg.
The timescale of V1065~Cen is the same as that of LV~Vul ($f_{\rm s}=1.0$
against LV~Vul).  The data of V1065~Cen are the same as those in Figure 56 
of \citet{hac16kb}; the original data are taken from the archive of AAVSO
and SMARTS \citep{wal12}.  In panel (b), we dereddened the colors 
of V1065~Cen with $E(B-V)=0.45$.  See the text for detail.
\label{v959_mon_lv_vul_v1065_cen_v_bv_ub_logscale_no3}}
\end{figure}

\section{Time-stretching Method of Nova Light Curves}
\label{v959_mon}

Novae show rich varieties in their optical/IR light curves, 
but there is a strong similarity in the main decline behavior. 
\citet{kat94h} theoretically explained the main trend of nova light curves,
i.e., a more massive WD shows a faster evolution in the decay phase of
light curve based on the optically thick wind theory.  \citet{hac06kb}
calculated free-free emission light curves of various WD masses and chemical
compositions, and found that these theoretical light curves overlap
with each other by time-stretch. 
They call this property the universal decline law.  
After that, these authors have analyzed a number of nova light
curves and showed that observed light curves basically decay 
along with the free-free model light curves  
\citep{hac10k, hac15k, hac16k, hac18k}.
Using this universal decline law and time-stretching method,
they have determined the WD masses for a number of novae
\citep[e.g.,][]{hac07k, hac09ka, hac14k, hac15k, hac16k, hac18k,
kat09hc, kat15sh}.

Figure \ref{v959_mon_v1668_cyg_lv_vul_v_bv_ub_logscale_no3}(a) 
shows the $V$, soft X-ray ($0.3-0.8$~keV), and hard X-ray ($0.8-10$~keV)
light curves of V959~Mon on a logarithmic timescale. 
We added the light curves of two classical novae, LV~Vul 1968\#1 
and V1668~Cyg 1978, that have the most similar decline 
among our 60 samples that we have ever analyzed.  
(We show two similar novae IV~Cep and V1065~Cen later in Figures
\ref{v959_mon_lv_vul_iv_cep_v_bv_ub_logscale_no3} and
\ref{v959_mon_lv_vul_v1065_cen_v_bv_ub_logscale_no3}, respectively.)

The light curve data of V959~Mon shows a good agreement with the later
phase of LV~Vul and V1668~Cyg.  Also the $(B-V)_0$ evolution is very
similar to the upper branch of LV~Vul.  These similarities strongly
suggest that V959~Mon follows the universal decline law, i.e., 
similar evolution to LV~Vul and V1668~Cyg even in the early phase.

In the present paper, we regard that V959~Mon follows the universal 
decline law from the early phase, where we have no data, until
the later nebular phase.
We did this reproduction of the early light curve of V959~Mon
in Figure \ref{v959_mon_v1668_cyg_lv_vul_v_bv_ub_logscale_no3}.
The V light curve decays along $t^{-3}$ after $250-300$ days of
the outburst.  This strongly indicates that optically thick winds
stopped on day $250-300$ \citep[see, e.g., Figure 25 of][]{hac16k}.
The slope of $t^{-3}$ usually begins after the nebular phase starts
in many novae.  Therefore, this nova shows normal decline of novae.

Although the overall decline features are very similar to V959~Mon,
LV~Vul and V1668~Cyg evolve faster than V959~Mon by a factor of 1.38. 
Therefore, these two nova light curves are time-stretched 
by $f_{\rm s}= 1.38$, i.e., they are shifted toward right 
by $\Delta \log t =\log f_{\rm s}= \log 1.38 = 0.14$
in the logarithmic time (see Section \ref{timescaling_factor}
for detail).  In Figure 
\ref{v959_mon_v1668_cyg_lv_vul_v_bv_ub_logscale_no3},
we assumed that the outburst day is $t_{\rm OB}=$JD~2456066.5
(UT 2012 May 19.0) as explained in Section \ref{outburst_day}. 
Thus, the optical light curve started from day 83.

\subsection{Timescaling Factor}
\label{timescaling_factor}
Now we show how to determine the timescaling (or time-stretching)
factor of $f_{\rm s}=1.38$.
We try to overlap these three novae as much as possible, 
by shifting in the vertical and horizontal directions.  
For the horizontal shift, we change $\Delta \log t$ by 0.01 or 0.02 
steps and check whether the light/color curves overlap with each other.
We select the best one by eye.  In Figure 
\ref{v959_mon_v1668_cyg_lv_vul_v_bv_ub_logscale_no3}(a), (b), and (c),
we finally select a factor of $\Delta \log t = \log 1.38 = 0.14$ both
for LV~Vul and V1668~Cyg.  The shift in the vertical direction $\Delta V$
is also obtained by changing $\Delta V$ by 0.1 or 0.2 mag steps and
select the best one by eye.  In the present case, we select
$\Delta V=1.6$ mag for LV~Vul, and $\Delta V=-1.1$ mag for V1668~Cyg.
This is the first step of the time-stretching method explained below.
Here, we regard V959~Mon as a target nova, and LV~Vul and V1668~Cyg
as template novae.  
It should be noted that there are two branches of $(B-V)_0$ color curve of
LV~Vul in Figure \ref{v959_mon_v1668_cyg_lv_vul_v_bv_ub_logscale_no3}(b)
after the nebular phase started and we fit the V959~Mon color curve
with the bluer (upper) branch of LV~Vul.  We explain the two separated
branches later in Section \ref{color-magnitude_diagram}.

We have determined the time-shift $\Delta \log t = \log f_{\rm s}$ of
LV Vul (open blue squares) against V959~Mon
by assuming that the 10 data points of LV Vul
(upper branch) overlap to those of V959 Mon 
in the $(B-V)_0$ color curves in Figure 
\ref{v959_mon_v1668_cyg_lv_vul_v_bv_ub_logscale_no3}(b).
We shift the color curve of LV Vul in steps of $\Delta \log t = 0.01$
or 0.02 and find best match by eye from the 10 data points
between $\log t = 2.0$ and 2.5.
Its allowance is about $\log f_{\rm s} = 0.14 \pm 0.05$ by eye.
We also estimated the error by a least square fit and obtained
$0.14\pm0.06$ in the time-shift.

The vertical fit in the $V$ magnitude is also determined by eye.
In the case of LV Vul and V959 Mon, we searched for a best fit
by changing the vertical shift in steps of 0.1 or 0.2 mag between
$\log t=1.95$ and 2.8 in Figure 
\ref{v959_mon_v1668_cyg_lv_vul_v_bv_ub_logscale_no3}(a).
Its allowance is about 0.1 mag by eye. 
We checked the error by a least square method (for steps of
0.1 mag vertical shift) and obtained $\Delta V=1.6\pm0.2$.
In this case, the difference between LV Vul and V959 Mon
is relatively large around the break ($\log t=2.3-2.4$) of the V light
curve.  Thus, the determination errors are about $\Delta \log t =\pm0.05$
and $\Delta V=\pm0.1$ or $\pm0.2$ mag unless $V$ and color data are
largely scattered.  This value is consistent with the eye fitting.
The least square fit implicitly assumes that all the data have equal
weight for deviation, which may not be real, because the data set
observed at different observatories may not be uniform.
Thus, in the later part of this paper, we use fitting by eye.

\subsection{Outburst Day}
\label{outburst_day}
We also explain how to determine the outburst day 
($t_{\rm OB}=$ JD 2456066.5).
The outburst day should be before the gamma-ray detection day
($t_{\rm OB}<$ JD~2456100.5).
We time-stretch the LV Vul and V1668 Cyg light/color curves with 
$f_{\rm s}=1.38$ on a linear timescale.  Then, we overlap
the $V$ light and $(B-V)_0$ color curves of LV~Vul and V1668~Cyg 
to those of V959~Mon.  We assumed the outburst day of V959 Mon
to be the same as that of LV Vul which is time-stretched with 
$f_{\rm s}=1.38$.  

It should be noted that the value of $t_{\rm OB}$ is slightly dependent
on the value of $\log f_{\rm s}$ and vice versa.  Starting from 
a trial value of $t_{\rm OB}<$ JD~2456100.5, we repeated 
twice the procedure of $\log f_{\rm s}$ and $t_{\rm OB}$
determination and confirmed the convergence of $t_{\rm OB}$ (and
$\log f_{\rm s}$).

\subsection{Time-stretched Light Curves}
\label{time-stretch_light_curve}
Figure \ref{v959_mon_v1668_cyg_lv_vul_v_bv_ub_logscale_no3}(a)
also shows theoretical free-free emission $V$ and blackbody
X-ray light curves of a $1.1~M_\odot$ WD with the envelope chemical 
composition of Ne nova 3 for V959~Mon (red lines), which is explained
in detail later in Section \ref{model_light_curve}.  We also add another
model light curve of a $0.98~M_\sun$ WD with the envelope chemical 
composition of CO nova 3 for V1668~Cyg (black lines) and LV~Vul, which is 
time-stretched by $\Delta \log t= \log f_{\rm s}= \log 1.38 = 0.14$.  

The magnitude of the two template novae, LV~Vul and V1668~Cyg,
decay along with these theoretical lines until day 90, 
when the two novae entered the nebular phase.  In the nebular phase, 
strong emission lines contribute to the $B$ and $V$ bands, which are not
included in our model light curves, so the observed $V$ magnitude deviates
much from, and decay more slowly than, the model $V$ light curve. 

After the optically thick winds stop, the observed $V$ magnitude decays
like the straight solid blue line of $F_\nu\propto t^{-3}$. 
That shows the trend of homologously expanding ejecta, i.e., 
the ejecta mass is constant \citep[see, e.g.,][]{woo97, hac06kb}.
In Figure 
\ref{v959_mon_v1668_cyg_lv_vul_v_bv_ub_logscale_no3}(a),
the V light curve of V959~Mon decays along $t^{-3}$ 
after day $\sim 250-300$.
This strongly indicates that optically thick winds stopped on day 
$\sim 250-300$ \citep[see, e.g., Figure 41 of][for such an example]{hac16k}.
We suppose that optically thick winds stopped on day $\sim 250-300$ for
V959~Mon.

We also plot two more novae that show similar decay timescale and shape
to V959~Mon, that is, IV~Cep 1971 and V1065~Cen 2007, in
Figures \ref{v959_mon_lv_vul_iv_cep_v_bv_ub_logscale_no3} and
\ref{v959_mon_lv_vul_v1065_cen_v_bv_ub_logscale_no3}, respectively.
The timescaling factor of these two novae are the same as that of 
LV~Vul ($f_{\rm s}=1$ against LV~Vul).
V1065~Cen show a dip in the $V$ light curve as shown in Figure 
\ref{v959_mon_lv_vul_v1065_cen_v_bv_ub_logscale_no3}.
This is because a dust shell formed \citep{hel10}.
Even though, the $V$ light curve show similar decay shape to that of
LV~Vul in the nebular phase.  IV~Cep and V1065~Cen follow the bluer branch
of the LV~Vul $(B-V)_0$ color curve as shown in Figures 
\ref{v959_mon_lv_vul_iv_cep_v_bv_ub_logscale_no3}(b) and
\ref{v959_mon_lv_vul_v1065_cen_v_bv_ub_logscale_no3}(b), respectively.
These figures mean that we cannot know in some detail
how the V959~Mon light curve
behaves in the early phase only from the information of the nebular phase
and later.  In other words, we do not know whether V959~Mon showed
a smooth decline like LV~Vul, dust dip (V1668~Cyg and V1065~Cen),
or wavy structure (IV~Cep). However, we may safely assume that
the main trend of the light curve is well reproduced by the 
model light curve.

\subsection{Time-stretching Factor and Distance Modulus}
Here, we obtain the distance modulus of this nova 
based on the time-stretching method of nova light curves. 
\citet{hac10k, hac15k, hac16k, hac18k} 
showed that, if the two nova $V$ light curves, i.e.,
one is called the template and the other is called the target,
$(m[t])_{V,\rm target}$ and $(m[t])_{V,\rm template}$
overlap each other after time-stretch by a factor of $f_{\rm s}$
in the horizontal direction and vertical shift by $\Delta V$, i.e.,
\begin{equation}
(m[t])_{V,\rm target} = \left((m[t \times f_{\rm s}])_V
+ \Delta V\right)_{\rm template},
\label{overlap_brigheness}
\end{equation}
their distance moduli in the $V$ band satisfy
\begin{equation}
(m-M)_{V,\rm target} = \left( (m-M)_V
+ \Delta V\right)_{\rm template} - 2.5 \log f_{\rm s}.
\label{distance_modulus_formula}
\end{equation}
Here, $(m-M)_{V, \rm target}$ and $(m-M)_{V, \rm template}$ are
the distance moduli in the $V$ band
of the target and template novae, respectively.
This is also written as
\begin{eqnarray}
\left( M'_V[t] \right)_{\rm target}
& \equiv & \left( M_V[t] \right)_{\rm target} - 2.5 \log f_{\rm s} \cr
& = & \left(M_V[t\times f_{\rm s}] \right)_{\rm template}
\label{distance-modulus_formula_prime_abs}.
\end{eqnarray}
This equation means that a target nova with a slower decline rate
($f_{\rm s} > 1$) than the template nova is fainter by 
$2.5 \times \log f_{\rm s}$, i.e., 
$M_V[t]_{\rm target} > M_V[t\times f_{\rm s}]_{\rm template}$. 

From Equation (\ref{distance_modulus_formula}),
we have the relation of
\begin{eqnarray}
(m&-&M)_{V, \rm V959~Mon}
= (m - M + \Delta V)_{V, \rm LV~Vul} - 2.5 \log 1.38 \cr
&=& 11.9\pm0.2 + 1.6\pm0.2 - 0.35 = 13.15\pm0.3 \cr
&=& (m - M + \Delta V)_{V, \rm V1668~Cyg} - 2.5 \log 1.38 \cr
&=& 14.6\pm0.2 - 1.1\pm0.2 - 0.35 = 13.15\pm0.3.
\label{distance_modulus_v959_mon_lv_vul_v1668_cyg}
\end{eqnarray}
Here, we take $f_{\rm s}=1.38$ and $\Delta V=+1.6\pm0.2$ for LV~Vul
and $f_{\rm s}=1.38$ and $\Delta V=-1.1\pm0.2$ for V1668~Cyg
as depicted in the figure ``LV~Vul V+1.6, 1.38 t'' 
and ``V1668~Cyg V$-1.1$, 1.38 t'' 
and adopt $(m-M)_{V, \rm LV~Vul}=11.9\pm0.2$ and 
$(m-M)_{V, \rm V1668~Cyg}=14.6\pm0.2$ from \citet{hac16kb}.
Thus, we obtain $(m-M)_V=13.15\pm0.3$ for V959~Mon.
From Equations (\ref{distance-modulus_formula_prime_abs})
and (\ref{distance_modulus_v959_mon_lv_vul_v1668_cyg}),
we have the relation of
\begin{eqnarray}
(m&-& M')_{V, \rm V959~Mon} 
= (m_V - (M_V - 2.5\log f_{\rm s}))_{\rm V959~Mon} \cr
&=& \left( (m-M)_V + \Delta V \right)_{\rm LV~Vul} \cr
&=& 11.9\pm0.2 + 1.6\pm0.2 = 13.5\pm0.3.
\label{absolute_mag_v959_mon_lv_vul}
\end{eqnarray}

We adopt the reddening of $E(B-V)=0.38$ after
Munari et al.'s (2013) value of $E(B-V)=0.38\pm0.01$ rather than
Shore et al.'s (2013) value of $E(B-V)=0.85\pm0.05$
(see discussion in Section \ref{discussion} for detail).
We also prefer to $E(B-V)=0.38$ because it gives a good matching
of the color curve to that of LV~Vul in Figure
\ref{v959_mon_v1668_cyg_lv_vul_v_bv_ub_logscale_no3}(b).
The distance is calculated from
\begin{equation}
(m-M)_V = 3.1 E(B-V) + 5 \log (d/10~{\rm pc}),
\label{distance_modulus_rv}
\end{equation}
where we adopt $R_V=A_V/E(B-V)=3.1$ \citep[e.g.,][]{rie85}.  We obtain
$d=2.5\pm0.5$~kpc together with $(m-M)_V=13.15\pm0.3$ 
and $E(B-V)=0.38\pm0.1$.
Table \ref{various_properties} summarizes these results.


\begin{deluxetable*}{llllllrl}
\tabletypesize{\scriptsize}
\tablecaption{Various Properties of V959~Mon and Selected Novae 
\label{various_properties}}
\tablewidth{0pt}
\tablehead{
\colhead{Object} & \colhead{Outburst} 
& \colhead{$\log f_{\rm s}$\tablenotemark{a}} 
& \colhead{$E(B-V)$} 
& \colhead{$(m-M)_V$} & \colhead{Distance} & \colhead{$z$\tablenotemark{b}} 
& \colhead{$M_{\rm WD}$} \\
\colhead{} & \colhead{Year} & \colhead{} & \colhead{} 
& \colhead{} & \colhead{(kpc)} 
& \colhead{(pc)} & \colhead{$(M_\sun)$} \\
}
\startdata
V1065~Cen & 2007 & 0.0 & 0.45 & 15.0 & 5.3 & 330 & 0.98\tablenotemark{c} \\
IV~Cep & 1971 & 0.0 & 0.65 & 14.5 & 3.1 & $-90$ & 0.98\tablenotemark{c} \\
V1668~Cyg\tablenotemark{d} 
& 1978 & 0.0 & 0.30 & 14.6 & 5.4 & $-635$ & 0.98\tablenotemark{c} \\
V1974~Cyg\tablenotemark{d} 
& 1992 & 0.03 & 0.30 & 12.2 & 1.8 & 245 & 0.98\tablenotemark{c} \\
V959~Mon & 2012 & 0.14 & 0.38 & 13.15 & 2.5 & 3 & 0.95\tablenotemark{c} \\
V959~Mon & 2012 & 0.14 & 0.38 & 13.15 & 2.5 & 3 & 1.05\tablenotemark{e} \\
V959~Mon & 2012 & 0.14 & 0.38 & 13.15 & 2.5 & 3 & 1.1\tablenotemark{f} \\
LV~Vul & 1968\#1 & 0.0 & 0.60 & 11.9 & 1.0 & 15 & 0.98\tablenotemark{c} \\
\enddata
\tablenotetext{a}{
$f_{\rm s}$ is the timescaling factor against LV~Vul.
}
\tablenotetext{b}{
$z$ is the distance from the galactic plane.
}
\tablenotetext{c}{
$M_{\rm WD}$ is obtained for the envelope chemical composition of
CO nova 3, i.e., $X=0.45$, $Y=0.18$, $Z=0.02$,
$X_{\rm CNO}=0.35$, $X_{\rm Ne}=0.0$ \citep{hac16k}.
}
\tablenotetext{d}{
Various parameters are taken from \citet{hac16k}.
}
\tablenotetext{e}{
$M_{\rm WD}$ is obtained for the envelope chemical composition of
Ne nova 2, i.e., $X=0.55$, $Y=0.30$, $Z=0.02$,
$X_{\rm CNO}=0.10$, $X_{\rm Ne}=0.03$ \citep{hac10k}.
}
\tablenotetext{f}{
$M_{\rm WD}$ is obtained for the envelope chemical composition of
Ne nova 3, i.e., $X=0.65$, $Y=0.27$, $Z=0.02$,
$X_{\rm CNO}=0.03$, $X_{\rm Ne}=0.03$ \citep{hac16k}.
}
\end{deluxetable*}

Figures \ref{v959_mon_lv_vul_iv_cep_v_bv_ub_logscale_no3} and
\ref{v959_mon_lv_vul_v1065_cen_v_bv_ub_logscale_no3} show
the light/color curves of IV~Cep and V1065~Cen in comparison with LV~Vul.
The timescales and $(B-V)_0$ color curve shapes of these two novae
are almost the same as those of LV~Vul and V1668~Cyg.  
Therefore, we apply Equations (\ref{overlap_brigheness}) 
and (\ref{distance_modulus_formula}) to Figure 
\ref{v959_mon_lv_vul_iv_cep_v_bv_ub_logscale_no3} and obtain the relation of
\begin{eqnarray}
(m&-&M)_{V, \rm IV~Cep}
= (m - M + \Delta V)_{V, \rm LV~Vul} - 2.5 \log 1.0 \cr
&=& 11.9\pm0.2 + (1.6\pm0.2 - (-1.0\pm0.2)) - 0.0 \cr
&=& 14.5\pm0.3.
\label{distance_modulus_iv_cep_lv_vul}
\end{eqnarray}
The value of $(m-M)_{V, \rm IV~Cep}=14.5\pm0.3$ is consistent
with the previous estimate of $(m-M)_{V, \rm IV~Cep}=14.7\pm0.2$
in \citet{hac16kb}.  
The distance is calculated to be $d=3.1\pm0.6$~kpc from Equation 
(\ref{distance_modulus_rv}) together with $E(B-V)=0.65\pm0.5$ \citep{hac16kb}.

In the same way, we obtain 
\begin{eqnarray}
(m&-&M)_{V, \rm V1065~Cen}
= (m - M + \Delta V)_{V, \rm LV~Vul} - 2.5 \log 1.0 \cr
&=& 11.9\pm0.2 + (1.6\pm0.2 - (-1.5\pm0.2)) - 0.0 \cr
&=& 15.0\pm0.3,
\label{distance_modulus_v1065_cen_lv_vul}
\end{eqnarray}
for V1065~Cen.
This value of $(m-M)_{V, \rm V1065~Cen}= 15.0\pm0.3$ is slightly
smaller than, but in reasonable agreement with the previous value of
$(m-M)_{V, \rm V1065~Cen}=15.3\pm0.2$ \citep{hac16kb}. 
The distance is calculated to be $d=5.3\pm1.0$~kpc
from Equation (\ref{distance_modulus_rv}) together with 
$E(B-V)=0.45\pm0.05$ \citep{hac16kb}.
These results are summarized in Table \ref{various_properties}.
Then, we have the relation of
\begin{eqnarray}
(m&-&M)_{V, \rm V959~Mon}
= (m - M + \Delta V)_{V, \rm IV~Cep} - 2.5 \log 1.38 \cr
&=& 14.5\pm0.3 -1.0\pm0.2 - 0.35 = 13.15\pm0.3 \cr
&=& (m - M + \Delta V)_{V, \rm V1065~Cen} - 2.5 \log 1.38 \cr
&=& 15.0\pm0.3 -1.5\pm0.2 - 0.35 \cr
&=& 13.15\pm0.3,
\label{distance_modulus_v959_mon_lv_vul_iv_cep_v1065_cen}
\end{eqnarray}
from Figures \ref{v959_mon_lv_vul_iv_cep_v_bv_ub_logscale_no3} and
\ref{v959_mon_lv_vul_v1065_cen_v_bv_ub_logscale_no3}.

In our method, there are two sources of ambiguity in $(m-M)_V$: one is the
$(m-M)_V$ error of the template nova and the other is the vertical 
$\Delta V$ fit error.  For the vertical fit, we change $\Delta V$
in steps of 0.1 mag and search for the best overlap by eye.
This error is typically 0.1 or 0.2 mag unless the $V$ data
are scattered.  The $(m-M)_V$ errors of templates are
dependent on each template (typically 0.2 mag).
We checked the fit with a least square method and obtained errors of
$\Delta V=\pm0.2$ or $\pm0.3$ mag.
Thus, the errors of distance modulus $(m-M)_V$ are 0.2 or 0.3 mag
unless otherwise specified.


\begin{figure}
\epsscale{1.15}
\plotone{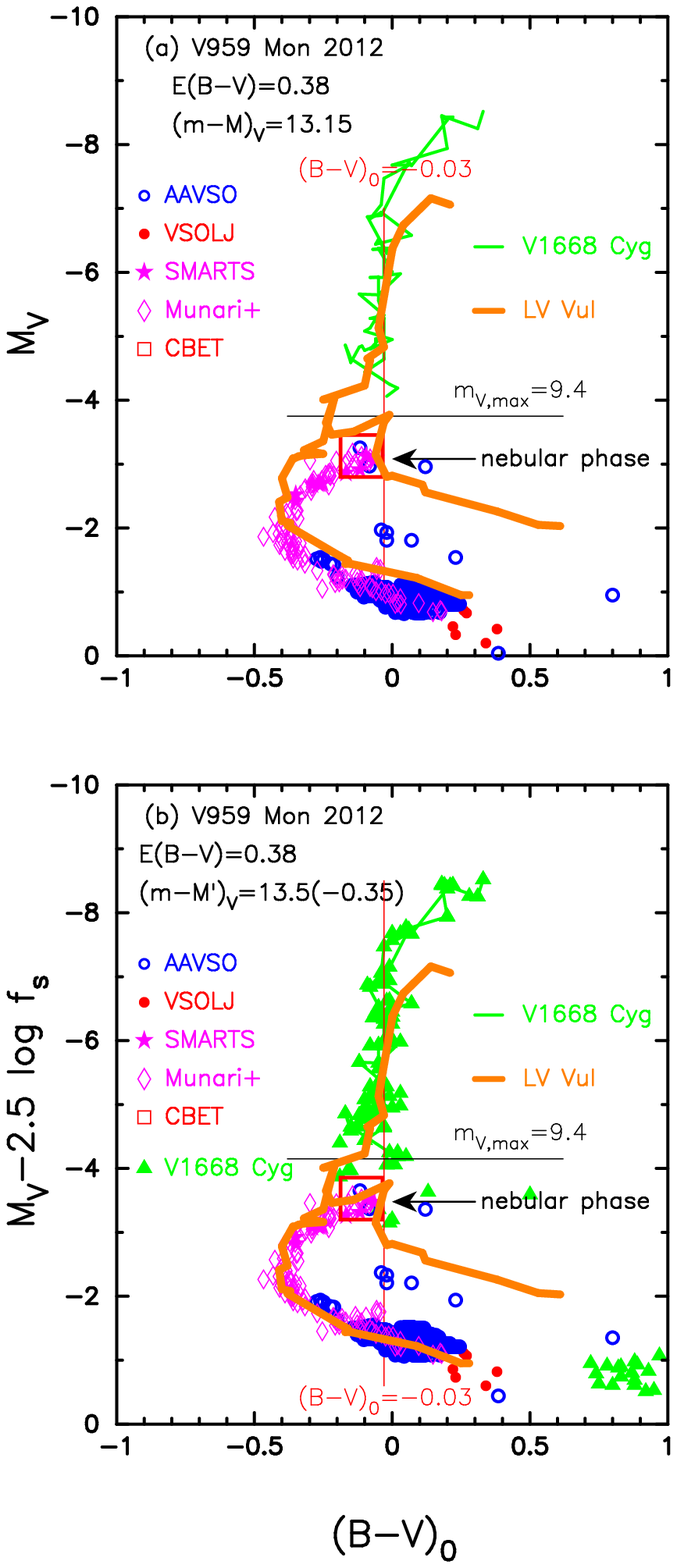}
\caption{
(a) Color-magnitude diagram of V959~Mon.
The ordinate is the absolute $V$ magnitude, $M_V$,
and the abscissa is the dereddened color, $(B-V)_0$.
The solid orange and green lines indicate the tracks
of LV~Vul and V1668~Cyg, respectively.  The vertical solid red line indicates
$(B-V)_0=-0.03$, the color of optically thick free-free emission.
The onset of nebular phase is indicated by a large open red square.
(b) Color-magnitude diagram of V959~Mon,
but the ordinate is the stretched absolute $V$ magnitude, 
$M'_V\equiv M_V-2.5\log f_{\rm s}$.  See the text for detail.
\label{hr_diagram_v959_mon_lv_vul_v1668_cyg_general_2fig}}
\end{figure}


\begin{figure}
\epsscale{1.15}
\plotone{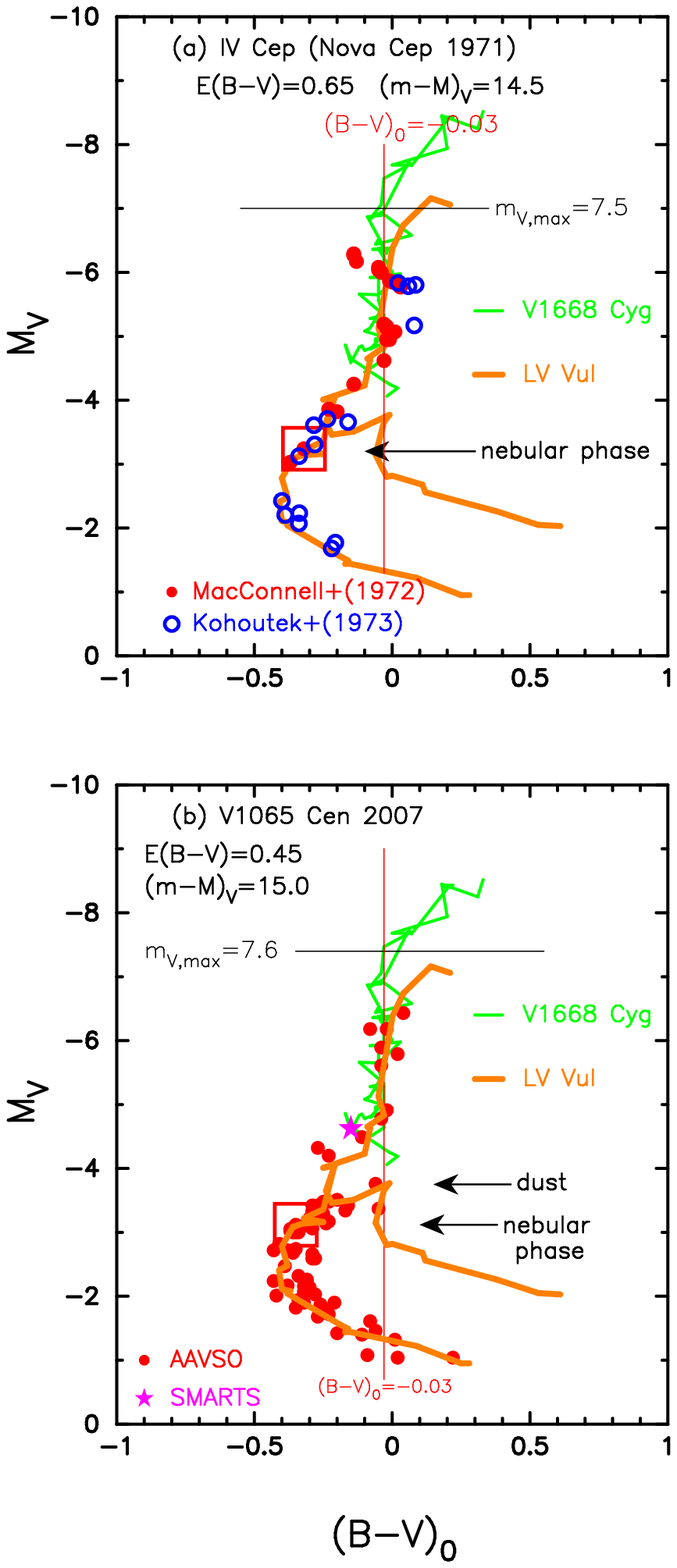}
\caption{
Color-magnitude diagrams of (a) IV~Cep and (b) V1065~Cen.
The ordinate is the absolute $V$ magnitude, $M_V$,
and the abscissa is the dereddened color, $(B-V)_0$.
The solid orange and green lines indicate the template tracks
of LV~Vul and V1668~Cyg, respectively.
The vertical solid red line indicates $(B-V)_0=-0.03$, 
the color of optically thick free-free emission.
The onset of nebular phase is indicated by a large open red square.
\label{hr_diagram_iv_cep_v1065_cen_2fig}}
\end{figure}

\section{Color-magnitude Diagram}
\label{color-magnitude_diagram}
\citet{hac16kb} analyzed 48 novae in the color-magnitude diagram, 
and showed that a typical nova evolves down along with 
the line of $(B-V)_0=-0.03$ in the early phase ($M_V < -4$) of the outburst. 
This indicates that the optical flux is dominated by free-free emission
because the intrinsic color of optically thick free-free
emission is $(B-V)_0=-0.03$ \citep{hac14k}.
The nebular phase begins when the nova becomes
as faint as $M_V \sim -4$ \citep[more exactly Equation (5) or (6) 
of][]{hac16kb}.  Figure 
\ref{hr_diagram_v959_mon_lv_vul_v1668_cyg_general_2fig}(a)
shows the outburst evolution of V1668 Cyg and LV~Vul 
in the color-magnitude diagram,
in which they go down along with the line of $(B-V)_0=-0.03$. 
In the nebular phase, the data of LV~Vul splits into two branches. 
This splitting is caused by slightly different response of V-band filters
\citep[see, e.g., Figure 1 of ][]{mun13b}.
Slightly different response functions of different $V$ filters produce
large differences in the $V$ magnitudes because of large
contributions of strong emission lines of [\ion{O}{3}] at the blue edge
of the $V$ filter \citep[see discussion of][]{hac06kb, hac14k, hac16kb}. 

In this figure we also plot the V959~Mon data for
$E(B-V)=0.38$ and $(m-M)_V=13.15$.  The data follow
the bluer branch of the LV~Vul track, but slightly fainter.  In the 
spectra of V959~Mon obtained by \citet{mun12} on UT 2012 August 20.14,
[\ion{O}{3}] 5007 line slightly exceeds that of H$\beta$.  
Therefore, we identify the start of 
the nebular phase at this day, i.e., $(B-V)_0=0.269 - 0.40= -0.13$ and 
$M_V=10.022 - 13.15= -3.128$,
which is denoted by the large open red square in Figure
\ref{hr_diagram_v959_mon_lv_vul_v1668_cyg_general_2fig}(a).

The evolution of V959~Mon in 
Figure \ref{hr_diagram_v959_mon_lv_vul_v1668_cyg_general_2fig}(a)  
does not exactly follow the LV~Vul evolution,  
whereas the color evolutions of these two well agree  
in Figure \ref{v959_mon_v1668_cyg_lv_vul_v_bv_ub_logscale_no3}(b). 
It is because we do not include time-stretching effect in 
Figure \ref{hr_diagram_v959_mon_lv_vul_v1668_cyg_general_2fig}(a). 
Here, we make another color-magnitude diagram, 
Figure \ref{hr_diagram_v959_mon_lv_vul_v1668_cyg_general_2fig}(b),
taking into account the time-stretching effect.
Remember that, in Figure 
\ref{v959_mon_v1668_cyg_lv_vul_v_bv_ub_logscale_no3},
the V959~Mon data agree well with that of LV~Vul only when we use
the stretched time and the stretched magnitude, $M'_V[t]_{\rm target}$
in Equation (\ref{distance-modulus_formula_prime_abs}). 
In the same way, we use $M'_V[t]_{\rm target}$ in the color-magnitude
diagram, instead of $M_V$, i.e., we shift the magnitude upward 
by $2.5 \times \log 1.38=0.35$ mag. 
The text ``$(m-M')_V=13.5(-0.35)$'' in the figure means that
$(m-M')_V=13.5$ and $(m-M)_V=13.5 - 0.35 = 13.15$.

Next, we obtain the intrinsic color $(B-V)_0$ with time-stretching effect. 
The stretched absolute $B$ and $V$ magnitudes of V959~Mon can be written 
against those of LV~Vul as,
\begin{eqnarray}
(M'_B[t])_{\rm V959~Mon} &\equiv& (M_B[t] - 2.5 \log f_{\rm s})_{\rm V959~Mon}
\cr
&=& (M_B[t\times f_{\rm s}])_{\rm LV~Vul}, 
\end{eqnarray}
and
\begin{eqnarray}
(M'_V[t])_{\rm V959~Mon} &\equiv& (M_V[t] - 2.5 \log f_{\rm s})_{\rm V959~Mon}
\cr
&=& (M_V[t\times f_{\rm s}])_{\rm LV~Vul}. 
\end{eqnarray}
Thus, we obtain
\begin{eqnarray}
(B-V)_{0, \rm V959~Mon}[t]&\equiv& (M_B[t] - M_V[t])_{\rm V959~Mon} \cr
&=& (M'_B[t] - M'_V[t])_{\rm V959~Mon} \cr
&=& (M_B[t\times f_{\rm s}] - M_V[t\times f_{\rm s}])_{\rm LV~Vul} \cr 
&\equiv& (B-V)_{0, \rm LV~Vul}[t \times f_{\rm s}]
\end{eqnarray}
This means that the intrinsic color is unchanged 
after the time-stretching process.  

Figure \ref{hr_diagram_v959_mon_lv_vul_v1668_cyg_general_2fig}(b) shows
that the resultant track of V959~Mon well overlaps with that of LV~Vul. 
The onset of the nebular phase of LV~Vul is close to that of V959~Mon
in the stretched color-magnitude diagram \citep[see Figure 6 of][]{hac16kb}.
Conversely, this agreement supports our values of $E(B-V)=0.38$,
$(m-M')_V=13.5$, $f_{\rm s}=1.38$. 
Thus, we confirm that the distance to V959~Mon is 
$d=2.5\pm0.5$~kpc from Equation (\ref{distance_modulus_rv})
together with $E(B-V)=0.38\pm0.1$ and $(m-M)_V=13.15\pm0.3$. 

We plot the color-magnitude diagrams of IV~Cep and V1065~Cen in Figure 
\ref{hr_diagram_iv_cep_v1065_cen_2fig}(a) and (b), respectively. 
The track of these two novae do not move after time-stretch
because the timescaling factor is $f_{\rm s}=1.0$ against that of LV~Vul.
These novae evolve down along with the bluer branch of LV~Vul.  
In IV Cep, strong emission lines of [\ion{O}{3}] appeared 
between UT 1971 September 12 and 22 \citep{ros75},
which is an indication of the nebular phase.
We identify the start of the nebular phase at $(B-V)_0=-0.37$ and 
$M_V=-3.23$, denoted by the large open red square in Figure 
\ref{hr_diagram_iv_cep_v1065_cen_2fig}(a).
This starting point is close to that of LV~Vul.
In V1065~Cen, we specify the starting point of dust blackout \citep{hel10}
at $(B-V)_0=-0.06$ and $M_V=-4.06$, denoted by the black arrow
in Figure
\ref{hr_diagram_iv_cep_v1065_cen_2fig}(b).
\citet{hel10} pointed out that the nova entered the early nebular
phase at $m_V\approx12$, about 70 days after maximum.  We denote
this phase by the large open red square, at $(B-V)_0=-0.35$
and $M_V=-3.42$.  This point is close to that of LV~Vul.

We list our results in Table \ref{various_properties}.
These results are consistent with each other and confirm that our
time-stretching method works among these nova systems.


\begin{figure}
\epsscale{1.15}
\plotone{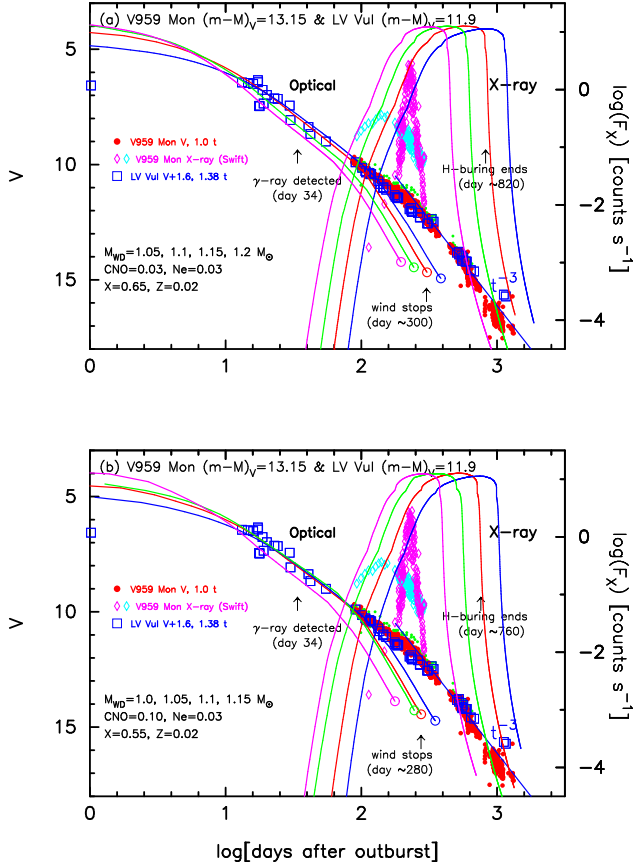}
\caption{
The model light curve fitting to V959~Mon. The filled red circles
and open blue squares represent the $V$ light curves of V959~Mon and
LV~Vul, respectively.  All observational data are the same as those in
Figure \ref{v959_mon_v1668_cyg_lv_vul_v_bv_ub_logscale_no3}(a). 
(a) The solid blue, red, green, and magenta lines
correspond to the $V$ and soft X-ray light curves of our 1.05, 1.1, 1.15,
and $1.2~M_\sun$ WDs, respectively, 
for the chemical composition of Ne nova3.
(b) The similar model light curve fitting of our 1.0 (blue), 1.05 (red),
1.1 (green), 
and $1.15~M_\sun$ (magenta) WDs for the chemical composition of Ne nova 2.
See the text for detail.
\label{v959_mon_mass_x65_x55_logscale_2fig}}
\end{figure}


\begin{figure}
\epsscale{1.2}
\plotone{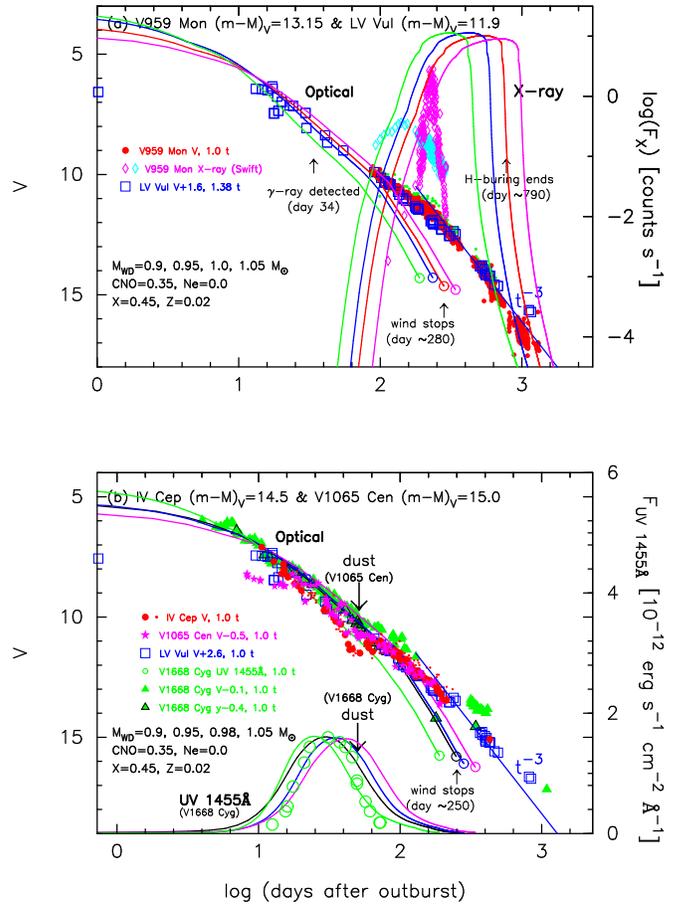}
\caption{
The model light curve fitting for the chemical composition
of CO nova 3 \citep{hac16k}. 
(a) All observational data are the same as those in
Figure \ref{v959_mon_mass_x65_x55_logscale_2fig}(a). 
The solid magenta, red, green, and blue lines
correspond to the $V$ and soft X-ray light curves of our 0.9, 0.95, 1.0,
and $1.05~M_\sun$ WDs, respectively.
(b) The similar model light curve fitting of our 0.9 (magenta),
0.95 (blue), 0.98 (black), and $1.15~M_\sun$ (green) WDs with
V1668~Cyg, LV~Vul, IV~Cep, and V1065~Cen.
See the text for detail.
\label{v959_mon_iv_cep_v1065_cen_mass_x45_x55_logscale_2fig}}
\end{figure}

\section{Model Light Curve Fitting}
\label{model_light_curve}
\subsection{WD masses of neon novae}
\label{neon_nova_wd_mass}
Neon novae are a subclass of classical novae that show neon emission lines
stronger than the permitted lines in the nebular phase.
Such neon enrichment is considered to originate from the core material
of an oxygen-neon (ONe) WD \citep[e.g.,][]{geh98}.
A natal ONe WD is likely massive
than $\gtrsim1.07~M_\sun$ \citep[e.g.,][]{ume99}, or $\gtrsim1.0~M_\sun$
\citep[e.g.,][]{wei00}, and has a thin helium-rich layer above 
a carbon--oxygen(CO)-rich mantle, e.g., a $0.035~M_\sun$ CO mantle for a
$1.09~M_\sun$ ONe core \citep{gil01}.  Such a WD may undergo
a number of nova explosions before the thin helium-rich layer is blown off.
Subsequently, the WD further undergoes a number of nova explosions before
the CO mantle on the ONe core is blown off.
If the mass accretion rate is $\sim 10^{-9} M_\sun$~yr$^{-1}$,
the ignition mass is $\sim 10^{-5} M_\sun$, and the same amount of
WD material is dredged up in every nova outburst, we can expect that 3,000
to 4,000 outbursts (in a total time of at least 30 -- 40 Myr)
must occur before the WD is deprived of its $0.035~M_\sun$
CO-rich mantle and significant neon is detected in the ejecta.
This could be a lower estimate because a CO-rich mantle
is more massive for the lower mass limit of the ONe core
\citep[as massive as $\sim0.1~M_\sun$,][]{gil03}.  Therefore,
the minimum masses of naked ONe cores
could be slightly smaller than $\sim 1.0~M_\sun$.

\subsection{Chemical composition of V959~Mon ejecta}
The largest ambiguity in the WD mass determination is in the choice of 
the chemical composition.  In many novae, chemical composition of ejecta
is not well determined. In our previous work \citep[e.g.,][]{hac16k}, we 
chose several sets of chemical composition as templates
(Ne nova 1, Ne nova 2, and Ne nova 3, which are basically calculated
from the degree of mixing between hydrogen-rich envelope and WD 
core material).  

\citet{sho13} obtained the chemical abundance of V959~Mon from
their optical spectra.  We converted their number-based abundance values
to our mass-weighted values, that is, 
$X=0.62$, $Y=0.25$, $X_{\rm C}=0.002$, $X_{\rm N}=0.026$, 
$X_{\rm O}=0.045$, $X_{\rm Ne}=0.059$, and $X_{\rm Fe}=0.001$ by weight.
\citet{tar14a} also estimated the chemical composition of V959~Mon from
her optical spectra, which are converted to 
$X=0.51$, $Y=0.29$, $X_{\rm N}=0.003$, $X_{\rm O}=0.006$, $X_{\rm Ne}=0.10$, 
and $X_{\rm Fe}=0.001$ by weight, assuming $X_{\rm C}=0.0$ because
she did not obtain the carbon abundance.
In the present work, we adopt Ne nova 3 
as a standard case because it is close to the abundance obtained by
\citet{sho13}, and Ne nova 2 and CO nova 3 for comparison,
to examine how the WD mass depends on the chemical composition.

\subsection{Multiwavelength light curves of V959~Mon}
Figure \ref{v959_mon_v1668_cyg_lv_vul_v_bv_ub_logscale_no3}(a)
shows the $V$, visual, and X-ray data of V959~Mon.
The X-ray data observed with {\it Swift}/XRT are divided into two bands,
i.e., $0.3-0.8$~keV (supersoft) and $0.8-10.0$~keV (hard) \citep{pag13}.  
In general, the main emitting wavelength region during a nova outburst
shifts from optical, UV, and to supersoft X-ray, because the WD photosphere
moves inward and the effective temperature rises with time. 
This figure also shows UV 1455 \AA\  flux of V1668 Cyg observed with 
{\it IUE}. We have no UV 1455 \AA\  observation for V959~Mon, but 
its UV phase should be the same. 
The supersoft X-ray flux of  V959~Mon increases after the UV dominant phase.
This is consistent with the general picture of nova evolutions that 
the main emitting wavelength region shifts to shorter one with time.  
Thus, we naturally regard that this supersoft X-ray is the emission
from the WD surface.  

\citet{hac10k} calculated a number of multiwavelength light curves with 
various WD masses and chemical compositions. The $V$ band fluxes are
dominated by free-free emission.  The UV~1455\AA\  and 
SSS fluxes are obtained assuming
blackbody emission at the photosphere.  The decay timescale of their 
model light curves depend mainly on the WD mass and weakly on the chemical
composition of the WD envelope.  Assuming the chemical composition of
Ne nova 3 \citep{hac06kb, hac10k, hac16k}, 
we calculated four model light curves of
1.05, 1.1, 1.15, and $1.2~M_\sun$ WDs and chose a best-fit model
of $1.1~M_\sun$ WD among these four model light curves as shown in
Figure \ref{v959_mon_mass_x65_x55_logscale_2fig}(a).
We list our results in Table \ref{various_properties}.
We discuss the model light curve fitting in detail
below in Section \ref{dependence_chemical_composition}. 
With this $1.1~M_\sun$ WD model and $(m-M)_V=13.15$ for V959~Mon,
we plot the model light curves
of absolute $V$ (left solid red line) and arbitrarily scaled
X-ray (right solid red line) in Figure 
\ref{v959_mon_v1668_cyg_lv_vul_v_bv_ub_logscale_no3}(a). 
The $V$ light curve of V959~Mon is reproduced with our model light curve
of $1.1~M_\sun$ WD until the early nebular phase.

\subsection{Dependence of WD mass on chemical composition}
\label{dependence_chemical_composition}
We examine the WD mass of V959~Mon in more detail,
because the model light curves depend weakly on the chemical 
composition \citep{hac06kb}.
Figure \ref{v959_mon_mass_x65_x55_logscale_2fig}(a) shows 
the model dependence of $V$ and supersoft X-ray light curve on the
WD mass for the chemical composition of Ne nova 3.
Because the early phase of $V$ light curve was not observed,
we use the stretched data of LV Vul.
The model light curves of 1.15 and $1.2~M_\sun$ WDs decay too early
and the supersoft X-ray light curve of $1.05~M_\sun$ rises too late
compared with the observation.  

The $t^{-3}$ decay of $V$ light curve started about day $250-300$
after the outburst as discussed in Section \ref{time-stretch_light_curve}.
This epoch corresponds roughly to the epoch when the optically thick
winds stopped.  The optically thick winds of 1.15 and $1.1~M_\sun$ WDs
stopped roughly on day 250 and 300, respectively.  
Thus, we select the $1.1~M_\sun$ WD
for the chemical composition of Ne nova 3.

Figure \ref{v959_mon_mass_x65_x55_logscale_2fig}(b)
depicts four light curves of 1.0, 1.05, 1.1, and $1.15~M_\sun$ WDs
for a different chemical composition of Ne nova 2 \citep{hac10k}.
The $V$ light curve of $1.15~M_\sun$ WD decays too early. 
The supersoft X-ray light curve rises too early in the $1.1~M_\sun$  
and too late in the $1.0~M_\sun$ compared with the observation. 
Thus, we select the $1.05~M_\sun$ WD
for the chemical composition of Ne nova 2, although the $1.1~M_\sun$ WD
cannot be rejected.

If we further decrease the hydrogen content by weight to $X=0.45$ 
\citep[CO nova 3, see][]{hac16k} from $X=0.55$ (Ne nova 2) 
and $X=0.65$ (Ne nova 3),
we similarly obtain a WD mass of $0.95~M_\sun$ as a best-fit one
among 0.9, 0.95, 1.0, and $1.05~M_\sun$ as shown in Figure
\ref{v959_mon_iv_cep_v1065_cen_mass_x45_x55_logscale_2fig}(a).
We summarize these results in Table \ref{various_properties}.

It should be noted that the neon content hardly affects the 
timescales of model nova light curves mainly because neon is not
included in the hydrogen burning (CNO-cycle) and does not much
contributes to the opacity.   
In other words, the WD mass depends weakly on the chemical
composition, especially for the hydrogen content $X$.

We obtain $1.1\pm0.05~M_\sun$ for $X=0.65$, $1.05\pm0.05~M_\sun$
for $X=0.55$, and $0.95\pm0.05~M_\sun$ for $X=0.45$, considering
the relatively large ambiguity of soft X-ray fit.
This kind of tendency of WD mass on $X$ 
was already discussed in \citet{hac07k}.
Our model light curve fitting gives WD masses between
$M_{\rm WD}=0.9-1.15~M_\sun$ for the chemical compositions of
$X=0.45-0.65$.
This result is consistent with the above discussion
on the minimum mass of naked ONe cores.

\subsection{Emergence of companion star}
Hard X-ray ($0.8-10$ keV) flux of V959~Mon was detected
by {\it Swift} \citep{pag13} about 85 days after
the outburst and reached maximum at day $\sim 140$ 
and then began to decrease until day $\sim 350$ 
(see Figure \ref{v959_mon_v1668_cyg_lv_vul_v_bv_ub_logscale_no3}(a)).
The origin of hard X-rays was interpreted as 
internal shocks \citep{fri87}
formed by collision between two ejecta shells \citep{muk01}, 
or the shock between nova winds (optically thick winds) and the companion
star \citep[see, e.g.,][]{hac05k, hac06kb, hac08kc}.
If it is the second case, the hard X-ray emergence should be
coincident with the emergence of the companion from the WD
photosphere because hard X-ray is probably absorbed deep inside
the nova photosphere.  

Using the $1.1~M_\sun$ WD model for Ne nova 3, 
we estimate the epoch when the companion emerges from
the nova envelope. The orbital period of $P_{\rm orb}= 0.296$~days
(7.1 hr) was derived by \citet{osb13} and \citet{mun13b} from the orbital
modulations of X-ray and optical light curves, respectively.
The mass of the donor star may be estimated
from Warner's (1995) empirical formula, i.e.,
\begin{equation}
{{M_2} \over {M_\sun}} \approx 0.065 \left({{P_{\rm orb}}
\over {\rm hours}}\right)^{5/4},
\mbox{~for~} 1.3 < {{P_{\rm orb}} \over {\rm hours}} < 9,
\label{warner_mass_formula}
\end{equation}
that gives $M_2 = 0.75 ~M_\sun$. 
With the WD mass $M_1 = 1.1 ~M_\sun$, 
the separation is calculated to be $a = 2.3 ~R_\sun$,  
the effective Roche lobe radius is 
$R_1^* =   0.95 ~R_\sun$ for the primary component (WD) 
and $R_2^* =   0.79 ~R_\sun$ for the secondary component 
(main-sequence star). 
Our theoretical $1.1~M_\sun$ WD model 
predicts that the photosphere of the nova envelope shrinks
to the orbital size, 
i.e., $R_{\rm ph} \approx a$, at $t_{\rm emerge} \approx 65$ days 
and further shrinks to $R_{\rm ph} \approx a - R_2^*$ 
(orbit minus the companion's radius, 
i.e., the companion entirely emerges from the WD envelope)   
at $t_{\rm emerge} \approx 80$ days.

In our model the companion emerges at $t_{\rm emerge} \approx 65-80$ days, 
which is roughly consistent with 
the hard X-ray detection at 85 days. 
The decay of the hard X-ray may be explained as the decrease of 
the wind mass-loss rate.  In our model, the wind mass-loss rate
monotonically decreases and stops at $t_{\rm wind} \sim 300$ days.
More exactly, the wind mass-loss rate quickly decreases after 
$t_{\rm break} \sim 100$ \citep[see][]{hac06kb}.
This is consistent with the hard X-ray behavior that peaks about 
$t \sim 140 $ days, and after that it monotonically decreased  
until $\sim 300$ days.  From these coincidences,
we suggest that the hard X-ray emission could originate from 
shock between the ejecta and companion. 

For the chemical composition of Ne nova 2 ($M_{\rm WD}=1.05~M_\sun$), 
the emerging time of the companion star from the WD photosphere is about
$t_{\rm emerge}=70$~days. 
For CO nova 3 ($M_{\rm WD}=0.95~M_\sun$), 
the emerging time is about $t_{\rm emerge}=80$~days.  
These values are still consistent with the above argument 
on hard X-ray emergence.

\subsection{Epoch at gamma-ray detection}
The gamma-ray was detected about 50 days before the hard X-ray count
rate increased.  Comparing with the theoretical light curve of
the $1.1~M_\sun$ WD (Ne nova 3), 
the photospheric radius of nova envelope is $R_{\rm ph}\sim 7~R_\sun$
($\sim 3$ times the separation) and the wind mass loss rate is
$\dot M_{\rm wind}= 3.5 \times 10^{-5}~M_\sun$~yr$^{-1}$ at the time
of gamma-ray detection.  This may be a clue for the emission mechanism
of gamma-ray.  For the chemical composition of Ne nova 2, 
we obtain a best-fit model of $M_{\rm WD}=1.05~M_\sun$.
The photospheric radius is $R_{\rm ph}\sim 5~R_\sun$,
and the wind mass-loss rate is $\dot M_{\rm wind}
\sim 3\times 10^{-5}~M_\sun$~yr$^{-1}$ at the gamma-ray detection (day 34).
For CO nova 3 ($M_{\rm WD}=0.95~M_\sun$), $R_{\rm ph}\sim 8~R_\sun$ 
and $\dot M_{\rm wind} \sim 4\times 10^{-5}~M_\sun$~yr$^{-1}$
at the gamma-ray detection.

\subsection{Short duration of supersoft X-ray phase}
Figure \ref{v959_mon_v1668_cyg_lv_vul_v_bv_ub_logscale_no3}(a)
also shows the model light curve of the supersoft X-ray flux. 
It rises before the optically thick wind stops.  
We suppose that in the wind phase 
the SSS flux may be partly blocked due to self-absorption by the wind.  
For example, in V1974~Cyg, only weak SSS flux was observed before the  
optically thick wind stopped, and after the wind stops, the supersoft
X-ray flux quickly increases \citep[see Figure 41 of][]{hac16k}. 
Considering these rising relations between the observed soft X-ray flux and
our model soft X-ray flux, the rise of the SSS flux in V959~Mon 
shortly before the wind stops is consistent with our model flux. 

The supersoft X-ray flux of V959~Mon decays much earlier than 
the model prediction, in which the X-ray turnoff time is calculated
from the end of hydrogen nuclear burning.  This observed short SSS duration 
($\sim 100$ days) in V959~Mon is much shorter than the SSS duration
of other classical novae with a similar timescale and WD mass. 
For example, V1974 Cyg shows the SSS duration of $\sim 350$ days
and its WD mass is estimated to be $0.98~M_\sun$ for the chemical
composition of CO nova 3 (and $f_{\rm s}=1.08$ against LV~Vul)
\citep[see, e.g.,][]{hac16k}. 

V959~Mon is a high inclination binary as mentioned in Section
\ref{introduction}.  The very short duration of the SSS phase
may be naturally explained with eclipse by a flaring-up rim of
the accretion disk.  In other words, the WD continues to emit 
supersoft X-rays from its surface until the end of the nuclear burning
but the WD surface was entirely obscured by a geometrically thick disk rim. 

\citet{hac03ka} presented a model to explain the light curve variation of 
the recurrent nova CI~Aql 2000 outburst by flaring-up of the disk rim. 
During the optically thick wind phase, the disk rim is partially blown off
in the wind, so the disk height at the rim is forced to remain small.
When the wind gradually weakens and finally stops, the disk rim flares up
because the wind shaping effect disappears.  This change of the disk rim
consistently explained the light curve variation of CI~Aql. 

As V959~Mon is a high inclination binary \citep[$i=82\arcdeg$:][]{rib13}, 
the WD and its surrounding emission region
could be perfectly shielded by the flaring-up disk rim when the wind stops 
\citep[see, e.g.,][for similar X-ray eclipses of V5116~Sgr]{sal08}.
This could cause the rapid decrease in the supersoft X-ray flux. 

It should be noted that in the case of V5116~Sgr 
($P_{\rm orb}=0.124$~days or 2.97 hr) the SSS emission is still
present during eclipse by the rim due to X-ray scattering in the
surrounding material \citep[e.g.,][]{sal17}.  The hardness of the X-ray
emission hardly changed during the eclipse of V5116~Sgr \citep{sal08}.  
On the other hand, if the flux decrease is due to a turnoff of 
hydrogen burning, one would expect the spectrum to become softer, colder, 
during the decline of the SSS.  For V959~Mon, \citet{pag13} reported 
no substantial change of X-ray spectra in the decline phase.
This may support our eclipse model by the disk rim.

\subsection{WD masses of LV~Vul, IV~Cep, and V1065~Cen}
The timescales of $V$ and color curves of LV~Vul,
IV~Cep, and V1065~Cen are very similar to that of V1668~Cyg. 
We regard that the WD masses of LV~Vul, IV~Cep, and V1065~Cen
are also similar to that of V1668~Cyg.  A comprehensive analysis of the
V1668~Cyg light curve was already presented in \citet{hac16k}, which
showed that the WD mass is $0.98\pm0.1~M_\sun$
for the chemical composition of CO nova 3.  We reproduce their
light curve analysis and plot four model light curves
for 0.9, 0.95, 0.98, and $1.05~M_\sun$ WDs in Figure
\ref{v959_mon_iv_cep_v1065_cen_mass_x45_x55_logscale_2fig}(b).


\begin{figure*}
\epsscale{1.0} 
\plotone{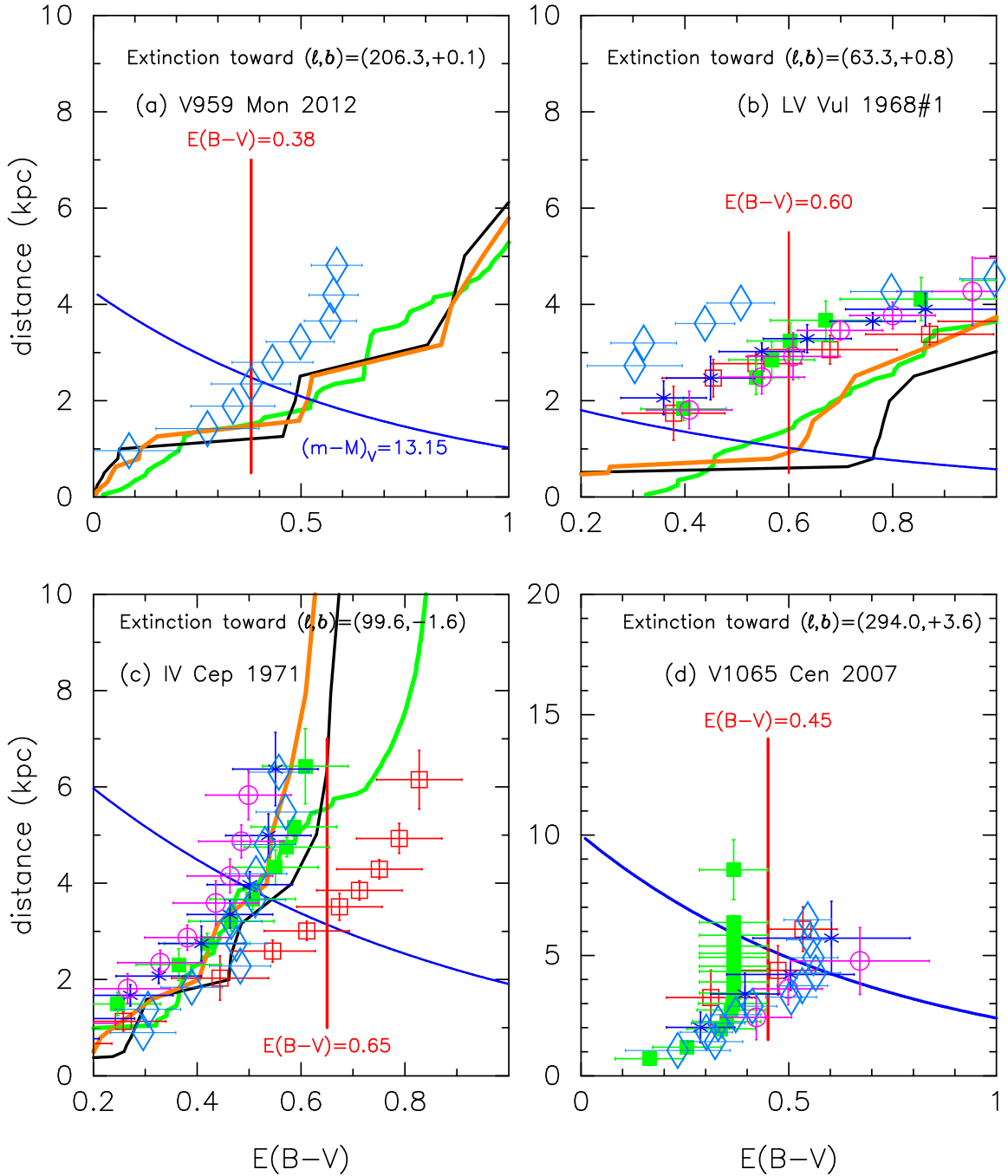}
\caption{
Distance-reddening relations toward (a) V959~Mon, (b) LV~Vul, (c) IV~Cep,
and (d) V1065~Cen.  The blue lines represent 
(a) $(m-M)_V=13.15$, (b) $(m-M)_V=11.9$, (c) $(m-M)_V=14.5$, 
and (d) $(m-M)_V=15.0$. 
The vertical red lines represent 
the color excess of each nova.  The black and orange lines represent
the distance-reddening relations given by \citet{gre15} and \citet{gre18},
respectively.  The green lines represent the relation given by \citet{sal14}.
The open cyan-blue diamonds with error bars represent the 
relation given by \citet{ozd16}.  The open red squares, filled green
squares, blue asterisks, and open magenta circles with error bars
represent the relations of \citet{mar06} in the four directions
close to each nova. 
\label{distance_reddening_v959_mon_lv_vul_iv_cep_v1065_cen_outburst}}
\end{figure*}

\section{Discussion}
\label{discussion}

\subsection{Reddening and distance}
\label{distance_discussion}
We examine the distances and reddenings toward the four novae
based on various distance-reddening relations.  
We use the following four results:
\citet{mar06} published a three-dimensional (3D) extinction map
of our galaxy in the direction of $-100\fdg0 \le l \le 100\fdg0$
and $-10\fdg0 \le b \le +10\fdg0$ with grids of $\Delta l=0\fdg25$
and $\Delta b=0\fdg25$.   \citet{sal14} calculated the reddening
for a region of $30\arcdeg \le l < 215\arcdeg$ and $|b|<5\arcdeg$
based on the IPHAS photometry.   
\citet{gre15} published data for the galactic extinction map,
which covers a wider range of the galactic coordinates 
(over three quarters of the sky) with much finer grids of 
3\farcm4 to 13\farcm7 and a maximum distance resolution of
25\%.  Their values of $E(B-V)$ could have an error of 0.05 -- 0.1 mag
compared with other two-dimensional (2D) dust extinction maps.
The distance-reddening relation was recently revised by \citet{gre18}.
\citet{ozd16} obtained distance-reddening relations toward 46 novae
based on the unique position of the red clump giants 
in the color-magnitude diagram.

\subsubsection{V959~Mon}
\label{V959_mon_distance_discussion}
Figure \ref{distance_reddening_v959_mon_lv_vul_iv_cep_v1065_cen_outburst}(a)
shows a few distance-reddening relations toward V959~Mon,
whose galactic coordinates are $(l,b)=(206\fdg3411, +0\fdg0758)$.
The reddening of $E(B-V)=0.38$ and the distance modulus of $(m-M)_V=13.15$
cross at $d=2.5$~kpc.  
The green line of \citet{sal14} is the nearest line-of-sight
reddening of $(l,b)=(206\fdg417, 0\fdg083)$.
The solid black and orange lines denote the distance-reddening relations
given by \citet{gre15} and \citet{gre18}, respectively.
The position of $d=2.5$~kpc and $E(B-V)=0.38$ is consistent with
that of \citet{ozd16} but slightly deviates from those of 
\citet{gre15, gre18} and \citet{sal14}.

\citet{mun13b} estimated the reddening to be $E(B-V)=0.38\pm0.01$
from the equivalent width of \ion{Na}{1} 5890 \AA\  line 
and derived the distance to be $\sim 1.5$~kpc by assuming that 
the companion is a K3 main-sequence star.
\citet{linf15} obtained the distance, $d=(0.9\pm0.2)$--$(2.2\pm0.4)$~kpc,
with a most probable distance of $d=1.4\pm0.4$~kpc, using the expansion
parallax method based on the VLA radio map.
On the other hand, \citet{sho13} derived a large 
reddening of $E(B-V)=0.85\pm0.05$ and the
hydrogen column density of $N_{\rm H}=(5\pm0.5)\times10^{21}$~cm$^{-2}$
by comparing the \ion{Na}{1} absorption line with
the Leiden/Argentine/Bonn (LAB) survey $\lambda$21~cm profile
of neutral hydrogen \citep{kal05}.  They also obtained the distance
of $3.6$~kpc from the direct comparison of UV and optical fluxes of
V959~Mon with those of V1974~Cyg, assuming that the distance and reddening
of V1974~Cyg are 3.6~kpc and $E(B-V)=0.36$.
\citet{tar14a} also derived the reddening of $E(B-V)=0.85$ from Balmer
decrements.  To summarize these previous studies, there are 
two distinct groups for the distance and reddening, 
one is $(d,E(B-V))=(1.5$~kpc$, 0.38)$ \citep[e.g.,][]{mun13b}
and the other is $(3.6$~kpc$, 0.85)$ \citep[e.g.,][]{sho13}.
The both sets of $(d,E(B-V))$ are close to the distance-reddening lines
obtained by \citet{gre15, gre18} and \citet{sal14}.

As mentioned earlier, \citet{linf15} obtained the distance of V959~Mon
based on the motion of the ejecta along the north-south axis.
The distance depends on the assumed expansion velocity of the ejecta
along the north-south axis.  Their distance values are between
the lower limit of $d=0.9\pm0.2$~kpc (assuming the expansion velocity
of $v_{\rm exp}=480\pm60$~km~s$^{-1}$) and the upper limit of 
$d=2.2\pm0.4$~kpc ($v_{\rm exp}=1200\pm150$~km~s$^{-1}$) from
the VLA radio map between day 126 and 199.  Their upper limit
is consistent with our distance of $d=2.5\pm0.5$~kpc.
\citet{linf15} also obtained ``a most probable distance of $d=1.4\pm0.4$~kpc,''
assuming $v_{\rm exp}=480\pm60$~km~s$^{-1}$ based on
the VLA map between day 615 and 703.  Note that their day 0 is defined
by UT 2012 June 19 (JD~2456097.5).  This $d=1.4$~kpc is not consistent with
$d=0.9\pm0.2$~kpc from the data between day 126 and 199 (using the same
$v_{\rm exp}=480\pm60$~km~s$^{-1}$).  
If we adopt $v_{\rm exp}=1200\pm150$~km~s$^{-1}$ instead,
their most probable distance increases to $d=3.5\pm1$~kpc for the data
between day 615 and 703.
Taking the arithmetic average of these two expansion velocities,
we obtain ``a most probable distance of (1.4+3.5)/2=2.45~kpc,'' 
which is close to our value of 2.5~kpc.
Thus, we may conclude that the distance of $2.5\pm0.5$~kpc is 
consistent with Linford et al.'s expansion parallax method.

\subsubsection{LV~Vul}
\label{lv_vul_distance_discussion}
For the reddening toward LV~Vul,
\citet{fer69} determined $E(B-V)=0.6\pm 0.2$ from the color excesses
of 14 B stars near the line of sight.
\citet{tem72} obtained $E(B-V)=0.55$ from the color at optical maximum;
i.e., $E(B-V)=(B-V)_{\rm max} - (B-V)_{0, \rm max}= 0.9 - 0.35 = 0.55$.
He adopted $(B-V)_{0, \rm max}= + 0.35$ \citep{sch57} instead of
$(B-V)_{0, \rm max}=+0.23$ \citep{van87}.  The distance toward LV~Vul
was obtained to be $d=0.92\pm 0.08$~kpc by \citet{sla95}
from the expansion parallax method.  
\citet{hac14k} obtained $E(B-V)=0.60\pm0.05$ by fitting with the 
typical color-color evolution track of nova outbursts.
These are all consistent with
our set of $d=1.0\pm0.2$~kpc and $E(B-V)=0.60\pm0.05$.
Thus, our set of $(d,E(B-V))$ for LV~Vul seems to be reasonable. 

  Figure 
\ref{distance_reddening_v959_mon_lv_vul_iv_cep_v1065_cen_outburst}(b)
shows several distance-reddening relations toward LV~Vul, 
$(l,b)=(63\fdg3024, +0\fdg8464)$.
We plot Marshall et al.'s distance-reddening relations of
four directions close to LV~Vul:
$(l,b)=(63\fdg25,0\fdg75)$ (red open squares),
$(l,b)=(63\fdg50,0\fdg75)$ (green filled squares),
$(l,b)=(63\fdg25,1\fdg00)$ (blue asterisks),
and $(l,b)=(63\fdg50,1\fdg00)$ (magenta open circles).
We added Green et al.'s, Sale et al.'s $(l,b)=(63\fdg250, 0\fdg917)$,
and \"Ozd\"ormez et al.'s distance-reddening relations.
In this way, various distance-reddening relations are not converged.

The large discrepancies among the distance-reddening relations
can be understood as follows.
The 3D dust maps essentially give an averaged value
of a relatively broad region, and thus the pinpoint reddening could be
different from the value of the 3D dust maps, because
the resolutions of these dust maps are considerably larger
than molecular cloud structures observed in the interstellar medium.
\citet{ozd16} used red clump giants. The number density of red clump giants
is smaller than that of giants that Marshall et al. used.
Therefore, the angular resolution of \citet{ozd16} could be less than
that of \citet{mar06}, although \citet{ozd16} claimed the accuracy
for the distance-reddening relation toward WY~Sge which is not significantly
different for the four resolutions of 0\fdg3, 0\fdg4, 0\fdg5, and 0\fdg8.
(WY~Sge is not included in the present analysis.)
The angular resolution of the map of Marshall et al. is 0\fdg25=15\farcm0.
Marshall et al. used only giants (or post main-sequence stars)
in their analysis, and thus the dust map
they produced has little information for the nearest kiloparsec.
Among these relations, only the orange line \citep{gre18}
is consistent with our 
estimates of $d=1.0\pm0.2$~kpc and $E(B-V)=0.60\pm0.05$.

\subsubsection{IV~Cep}
\label{iv_cep_distance_discussion}
The reddening toward IV~Cep was estimated to be $E(B-V)= 0.8$
\citep{sat73} from the interstellar absorption in the Cepheus region,
and $E(B-V)=A_V/3.1 = 1.8/3.1 = 0.58$ \citep{tho73} and  $E(B-V)=A_V/3.1 =
1.7/3.1 = 0.55$ \citep{koh73}, both from the absorption--distance
relation given by \citet{nec67}.  
\citet{hac16kb} obtained $E(B-V)=0.65\pm0.05$ by fitting with the 
typical color-color evolution track of nova outbursts.
All these values are roughly consistent with our estimate of
$E(B-V)=0.65\pm0.05$.  

Figure 
\ref{distance_reddening_v959_mon_lv_vul_iv_cep_v1065_cen_outburst}(c)
shows several distance-reddening relations toward IV~Cep, 
We plot four relations given by \citet{mar06} in directions
close to IV~Cep:
$(l, b)=(99\fdg5,-1\fdg5)$ (red open squares),
$(99\fdg75,-1\fdg5)$ (green filled squares),
$(99\fdg5,-1\fdg75)$ (blue asterisks),
and $(99\fdg75,-1\fdg75)$ (magenta open circles).
We also add Green et al.'s, Sale et al.'s $(l,b)=(99\fdg583, -1\fdg583)$,
and \"Ozd\"ormez et al.'s relations.  Among these relations, 
Marshall et al.'s relation of open red squares is consistent with our set of
$d=3.1\pm0.6$~kpc and $E(B-V)=0.65\pm0.05$.

\subsubsection{V1065~Cen}
\label{v1065_cen_distance_discussion}
Figure 
\ref{distance_reddening_v959_mon_lv_vul_iv_cep_v1065_cen_outburst}(d)
shows various distance-reddening relations toward V1065~Cen, 
$(l,b)=(293\fdg9836, +3\fdg6129)$.
We plot \"Ozd\"ormez et al.'s relation,  
and four distance-reddening relations given by \citet{mar06}:
$(293\fdg75,3\fdg75)$ by red open squares,
$(294\fdg00,3\fdg75)$ by filled green squares,
$(293\fdg75,3\fdg50)$ by blue asterisks, and
$(294\fdg00,3\fdg50)$ by open magenta circles.
The closest ones of \citet{mar06} 
are those denoted by filled green squares and open magenta circles.  
Our values of $d=5.3\pm1.0$~kpc and $E(B-V)=0.45\pm0.5$ are
midway between them. 

The reddening for V1065~Cen was obtained as $E(B-V)=0.50\pm0.10$ by
\citet{hel10} from an average of three estimates, i.e.,
$E(B-V)= (B-V)_{\rm max} - (B-V)_{0, \rm max} =
0.52\pm0.04 - (0.23 \pm 0.06) = 0.29\pm 0.07$,
$E(B-V)= (B-V)_{t2} - (B-V)_{0, t2} = 0.41\pm0.05 - (-0.02\pm 0.04)
= 0.43\pm 0.06$, $E(B-V)=0.79\pm0.01$ from the Balmer decrement
(H$\alpha$/H$\beta$).
\citet{hel10} also estimated the apparent distance modulus in the $V$ band
as $(m-M)_V= 7.6\pm0.2 - (-8.6\pm0.5)= 16.2\pm0.6$ from the MMRD
relation together with $t_2=11$ days.  This gives a distance of
$d=8.7^{+2.8}_{-2.1}$~kpc.  Note, however, that the distance estimate by
the MMRD relation is not so accurate \citep[see, e.g.,][]{dow00}.
\citet{hac16kb} obtained $E(B-V)=0.45\pm0.05$, assuming that the 
intrinsic $(B-V)_0$ color evolution of V1065~Cen is identical with 
that for similar types of novae, i.e., LV~Vul and V1668~Cyg.
This value is consistent with the estimate of Helton et al.'s value.

To summarize, our obtained distances and reddenings for the four novae
are broadly consistent with other estimates.  
We conclude that the distance of V959~Mon is $d=2.5\pm0.5$~kpc 
for $E(B-V)=0.38\pm0.01$.

\section{Conclusions}
\label{conclusions}

Our main results are summarized as follows:
\begin{enumerate}

\item The $V$ light curves of V959~Mon and LV~Vul overlap each other,
if we stretch the timescale of LV~Vul by a factor of $f_{\rm s}=1.38$.
Applying the time-stretching method to the $V$ light curves
of V959~Mon and LV~Vul, we obtain the distance modulus of V959~Mon
in the $V$ band, $\mu_V\equiv (m-M)_V=13.15\pm0.3$.  
The distance is calculated to be
$d=2.5\pm0.5$~kpc for the reddening of $E(B-V)=0.38\pm0.01$. 
We also apply the time-stretching method for other sets of V959~Mon vs.
V1668~Cyg, IV~Cep, and V1065~Cen, and obtain similar values
for the distance modulus $(m-M)_V$ for V959~Mon.

\item The stretched color-magnitude track of V959~Mon just overlaps with
that of LV~Vul in the $(B-V)_0$-$(M_V-2.5\log f_{\rm s})$ diagram.  
This strongly supports our adopted values of $E(B-V)=0.38\pm0.1$ 
and $(m-M')_V= (m-(M_V-2.5\log f_{\rm s}))=13.5\pm0.3$. 
The color-magnitude tracks of IV~Cep and V1065~Cen also overlap 
with that of LV~Vul, which may indicate a common track 
in the stretched color-magnitude diagram.

\item The various distance-reddening relations toward V959~Mon
are consistent with our obtained values.
Thus, we confirm $d=2.5\pm0.5$~kpc, $E(B-V)=0.38\pm0.1$,
and $(m-M)_V=13.15\pm0.3$. 

\item The model light curve fitting suggests the WD mass of
$M_{\rm WD}=0.9-1.15~M_\sun$, depending weakly on the assumed
chemical composition.  This range of the WD mass is 
consistent with the claim that V959~Mon is an ONe nova.
If we adopt the model chemical composition of Ne nova 3, which is close 
to the abundance obtained by \citet{sho13}, the WD mass is 
$M_{\rm WD}=1.1\pm0.05~M_\sun$.

\item The period of hard X-ray emission is consistent with the time
of appearance of the companion star from the nova envelope. 
Thus, the hard X-rays could be shock-origin between the ejecta
and companion. 

\item 
The supersoft X-ray flux increases in the later phase of a nova outburst
when the extended WD envelope becomes transparent to soft X-rays. 
The X-ray turn-on time of V959~Mon is quite consistent with 
theoretical model, which is confirmed by a number of classical novae. 
In other words, no diagnostics can be found for the X-ray turn-on time. 
This is a typical event of a nova evolution. 

\item
The very short supersoft X-ray phase in V959~Mon ($ \sim 100$ days) 
can be explained as the occultation by the disk rim. 
Our theoretical model indicates that the WD radius becomes as small 
as $R_{\rm ph}=0.08~R_\sun$ at the X-ray turnoff time that can be 
easily hidden by the disk rim ($i=82\arcdeg$). 
In short, the X-ray turnoff is due to the occultation by the disk rim,
instead of nuclear burning turnoff.

\item Our WD models suggest that, at the time of gamma-ray detection,
the photosphere of nova envelope
extends to $5-8~R_\sun$ (about two or three times the binary separation)
and the wind mass-loss rate is $(3-4)\times 10^{-5} M_\sun$~yr$^{-1}$.  
This may be a clue for the emission mechanism of gamma-ray.

\end{enumerate}

\acknowledgments
     We thank
the American Association of Variable Star Observers
(AAVSO) and the Variable Star Observers League of Japan (VSOLJ)
for the archival data of V959~Mon.
  We are also grateful to the anonymous referee for useful comments
 regarding how to improve the manuscript.
This research has been supported in part by Grants-in-Aid for
Scientific Research (15K05026, 16K05289)
from the Japan Society for the Promotion of Science.

\clearpage












\end{document}